\documentclass[11pt,a4paper]{article}
\usepackage{jheppub}
\usepackage{bera}
\usepackage[T1]{fontenc}
\usepackage[latin9]{inputenc}
\usepackage{color}
\usepackage{babel}
\usepackage{prettyref}
\usepackage{float}
\usepackage{units}
\usepackage{mathrsfs}
\usepackage{amsmath}
\usepackage{amssymb}
\usepackage{setspace}
\makeatletter

\@ifundefined{date}{}{\date{}}
\usepackage{upgreek}
\usepackage{tikz}
\newrefformat{fig}{Fig.\ref{#1}}
\newrefformat{tab}{Table \ref{#1}}
\newrefformat{sec}{Section \ref{#1}}

\makeatother

\title{Deconstruction hierarchies and locality diagrams of conformal models}

\author{P. Bantay}

\emailAdd{bantay@general.elte.hu}

\affiliation{Institute for Theoretical Physics, E{\" o}tv{\" o}s Lor{\' a}nd University, Budapest}
\abstract{The relationship between locality graphs and deconstruction hierarchies
of conformal models is explained, leading to computationally effective
procedures for determining the latter, and the relevant notions are
illustrated with several examples.}
\arxivnumber{2104.04964}

\begin{document}
\maketitle 

\global\long\def\set#1#2{\left\{  #1\, |\, #2\right\}  }%

\global\long\def\mat#1#2#3#4{\left(\begin{array}{cc}
#1 & #2\\
#3 & #4
\end{array}\right)}%

\global\long\def\Mod#1#2#3{#1\equiv#2\, \left(\mathrm{mod}\, \, #3\right)}%

\global\long\def\inv{^{\,\textrm{-}1}}%

\global\long\def\pd#1{#1^{+}}%

\global\long\def\sym#1{\mathbb{S}_{#1}}%

\global\long\def\fix#1{\mathtt{Fix}\!\left(#1\right)}%

\global\long\def\map#1#2#3{#1\!:\!#2\!\rightarrow\!#3}%

\global\long\def\Map#1#2#3#4#5{\begin{split}#1:#2  &  \rightarrow#3\\
 #4  &  \mapsto#5 
\end{split}
 }%

\global\long\def\fact#1#2{#1\slash#2}%

\global\long\def\zn#1{\left(\mathbb{Z}/\!#1\mathbb{Z}\right)^{\times}}%

\global\long\def\sn#1{\mathbb{S}_{#1}}%

\global\long\def\aut#1{\mathrm{Aut\mathit{\left(#1\right)}}}%

\global\long\def\FA#1{\vert#1\vert}%

\global\long\def\FI#1{#1_{{\scriptscriptstyle \pm}}}%

\global\long\def\qd#1{\mathtt{d}_{#1}}%

\global\long\def\cw#1{\mathtt{h}_{#1}}%

\global\long\def\om#1{\omega\!\left(#1\right)}%

\global\long\def\fc{\textrm{FC set}}%

\global\long\def\spr{\textrm{spread}}%

\global\long\def\sp#1{\boldsymbol{\upmu}\!\left(#1\right)}%

\global\long\def\v{\mathtt{0}}%

\global\long\def\fm{\mathtt{N}}%

\global\long\def\du#1{#1^{{\scriptscriptstyle \perp}}}%

\global\long\def\lat{\mathcal{\mathscr{L}}}%

\global\long\def\vera{\mathcal{V}}%

\global\long\def\svera#1{\vera_{#1}}%

\global\long\def\zquot#1#2{\mathfrak{#1}/\!#2}%

\global\long\def\vfc{\mathfrak{o}}%

\global\long\def\sc{\mathcal{\cov\vfc}}%

\global\long\def\lgr{\mathcal{G}}%

\global\long\def\defl#1{#1^{\prime}}%

\global\long\def\fcs{\textrm{FC set}}%
\global\long\def\pcl{\textrm{equilocality class}}%

\global\long\def\cl#1{\lgr^{2}\!\left(#1\right)}%

\global\long\def\latg#1{\mathbb{L}\!\left(#1\right)}%

\global\long\def\star#1{\lgr\!\left(#1\right)}%

\global\long\def\ve{\mathcal{V}}%

\section{Introduction}

Orbifold deconstruction \citep{Bantay2019a,Bantay2020}, the procedure
aimed at recognizing whether a given conformal model is an orbifold
\citep{Dixon_orbifolds1,Dixon_orbifolds2} of another one, and if
so, to determine both this original model and the relevant twist group,
is an interesting tool for the study of 2D CFT and, in a broader context,
of discrete gauge symmetries. Indeed, since orbi\-folding amounts
to gauging global symmetries of a conformal model \citep{DV3}, deconstructing
the orbifold sheds light on the effect of gauging selected symmetries.
This is all the more true if one considers the full hierarchy of deconstructions
that results from the possibility of orbi\-folding by stages: gauging
the global symmetries forming a group $G$ may be achieved in steps,
by first gauging the symmetries from a normal subgroup $N\!\triangleleft\!G$,
and then gauging the resulting model by the factor group $\fact GN$.
As a consequence, each orbifold model will have different partial
deconstructions corresponding to the different normal subgroups of
its twist group, and the hierarchy of these partial deconstructions
will be described by the lattice of normal subgroups of the latter.
But having such a hierarchy of partial deconstructions means that
we can study the effect of gauging layer by layer, leading to a more
refined understanding of the whole process. This idea points clearly
at the importance of controlling the full deconstruction hierarchy
for any given conformal model.

Actually, the situation is a bit more complicated, because in many
cases one and the same conformal model might be realized as an orbifold
in several fundamentally different ways. This means that there might
exist several maximal deconstructions of a given model \citep{Bantay2019a,Bantay2020},
each realizing it as an orbifold with a possibly different twist group
and/or deconstructed model: this phenomenon is already apparent for
the simplest case of $\mathbb{Z}_{2}$-orbifolds, e.g. in the construction
of the Moonshine module from the Leech lattice VOA \citep{FLM1}.
Having different maximal deconstructions means that the hierarchy
of all deconstructions of a given model cannot be described by a lattice,
but only by some more general ordered structure, which still has pretty
special features, like having all its order ideals isomorphic to the
normal subgroup lattice of some finite group.

The way out, as explained in \citep{Bantay2020a}, is to consider
the hierarchy of deconstructions as embedded in the lattice (ordered
by inclusion) of so-called $\fc$s, i.e. sets of primary fields closed
under the fusion product. That such an embedding is possible follows
from the observation that in any orbifold model there is a distinguished
set of primaries, the so-called vacuum block, which consists of those
primaries of the orbifold (in one-to-one correspondence with the irreducible
representations of the twist group) that originate in the vacuum sector
of the original model. This vacuum block has very special properties,
since it is closed under the fusion product, and all its elements
have integer conformal weight and quantum dimension, i.e. it is a
so-called twister \citep{Bantay2019a,Bantay2020}. Each twister of
a given model corresponds to a different deconstruction, realizing
it as an orbifold in a different way, and the hierarchy of deconstructions
is reflected by the inclusion relation among the twisters. Most importantly
for us, the basic features of the deconstruction hierarchy follow
directly from the fact that the lattice $\lat$ of $\fc$s is a modular
lattice endowed with an involutive and order reversing self-map \citep{Bantay2020a}.

It should transpire from the above that the major task for understanding
the deconstruction hierarchy of a given model is to determine the
corresponding lattice $\lat$ and the precise location of the different
twisters inside it. This is by no means a trivial job, since a brute
force approach would have a computational cost growing exponentially
with the number of primaries, and it would fail already for models
with 20-30 primaries, while really interesting examples usually involve
at least hundreds, if not thousands of them. At first sight this could
seem to be a major obstacle, but, as we shall explain, there is a
way out, exploiting the connection between the lattice $\lat$ and
the locality graph of the model. Not only does this give us an effective
procedure to determine $\lat$, but it does also explain some striking
features of the deconstruction hierarchy that could seem accidental
otherwise.

In the next section, we shall recall those results about the lattice
$\lat$ that are necessary for understanding the rest of the paper.
Then we shall turn to the relation between locality graphs and the
lattice of $\fcs$s, and explain how this sheds light on some properties
of deconstruction hierarchies not understood before, in particular,
why all but two Virasoro minimal models share the same lattice. Finally,
we shall comment on possible consequences and applications of the
results presented.

\section{FC sets and their lattice\label{sec:FC-sets-and}}

A set $\mathfrak{g}$ of primaries of a conformal model is fusion
closed, or an $\mathcal{\fc}$ for short \citep{Bantay2020a}, if
it contains the vacuum primary, and if the fusion product of any two
of its elements contains only primaries from $\mathfrak{g}$. In other
words, if $p$ and $q$ are elements of an $\mathcal{\fc}$ $\mathfrak{g}$,
and $N_{pq}^{r}\!>\!0$ for some primary $r$, then $r$ is also an
element of $\mathfrak{g}$. Note that an $\fc$ contains automatically
the charge conjugate of all its elements. The collection $\lat$ of
$\fcs$s of a given conformal model is partially ordered by inclusion,
with minimal element the trivial $\fcs$ that consists of the vacuum
solely, and maximal element the set of all primaries. Since the intersection
of two $\fcs$s is clearly an $\fcs$ again, $\lat$ is actually a
lattice \citep{Gratzer2011}.

As it turns out, $\lat$ is a lattice of a very special kind \citep{Bantay2020a}:
it is a self-dual lattice admitting a type I embedding into a partition
lattice \citep{Jonsson1953}, and in particular, it is a modular lattice
\citep{Gratzer2011}. This last property is fundamental in relation with orbifold deconstruction, since it guarantees that the set of partial
deconstructions of an orbifold is isomorphic to the normal subgroup
lattice of the twist group. Self-duality of $\lat$ refers to the
fact that to each $\mathfrak{g}\!\in\!\lat$ one can associate its
trivial class $\du{\mathfrak{g}}$, the collection of all those primaries
that are mutually local with every primary of $\mathfrak{g}$, which
is itself an $\fcs$, and the assignment $\mathfrak{g}\!\mapsto\!\du{\mathfrak{g}}$
is an involutive and order-reversing map of $\lat$ onto itself. If
all  elements of an $\fc$ $\mathfrak{g}$ have integer conformal
weight, i.e.  $\mathfrak{g}$ is a twister, there is a corresponding
realization of the conformal model as an orbifold, and the trivial
class $\du{\mathfrak{g}}$ consists of the primaries contained in
the untwisted sector. 

Local $\fcs$s, i.e. those  contained in their trivial class,
play a special role \citep{Bantay2020a}, for one may show that all local
$\fcs$s are either twisters, in which case they provide a realization
of the given model as an orbifold, or $\mathbb{Z}_{2}$-extensions
of twisters, in which case a suitable generalization of the deconstruction
procedure leads to a fermionic extension a la Runkel-Watts \citep{Runkel2020}.

An interesting class of $\fcs$s is formed by those whose elements
have integer quantum dimension. Such integral $\fcs$s include the
local ones \citep{Bantay2020a}, and form a sublattice of $\lat$,
with maximal element the set of all those primaries whose quantum dimension
is an integer. An interesting aspect of integral $\fcs$s is that
it is possible to associate to them a 'character table' which shares
many non-trivial properties of character tables of finite groups.
This is no surprise for twisters, since these correspond to some orbifold
realization of the model, and the resulting table is just the ordinary
character table of the twist group, but many deep analogies persist
even in cases when one can explicitly show (by excluding case-by-case all potential
candidates) that there is no suitable group with corresponding character table. These analogies
allow to generalize to integral $\fcs$s many notions from group theory, like  nilpotency and solubility \citep{Bantay2020a}, and suggest that
integral $\fcs$s might be related to some kind of 'generalized group
structure'.

There is one more important property of the lattice $\lat$ that should
be mentioned. To any collection $X$ of primary fields one can associate
 the sum
\begin{equation}
\sp X=\sum_{p\in X}\qd p^{2}\label{eq:sprdef}
\end{equation} 
where $\qd p$ denotes the quantum dimension of
the primary $p$. 
When $\mathfrak{g}\!\in\!\lat$ happens to be a twister corresponding
to a realization  of the conformal 
 model as an orbifold, $\sp{\mathfrak{g}}$
equals the order of the 
 relevant twist group \citep{Bantay2019a}. It is pretty
clear that
\begin{equation}
\sp{X\!\cup\!Y}\!+\!\sp{X\!\cap\!Y}\!=\!\sp X\!+\!\sp Y\label{eq:sprmod}
\end{equation}

\noindent and one has $\sp X\!\leq\!\sp Y$ for a subset $X\!\subseteq\!Y$.
The point is that  the product $\sp{\mathfrak{g}}\!\sp{\du{\mathfrak{g}}}$
is the same \citep{Bantay2020a} for all $\mathfrak{g}\!\in\!\lat$. As we shall see later, this 
leads to non-trivial restrictions that exclude some possibilities
that would look completely healthy otherwise.

\section{The locality diagram of a conformal model\label{sec:The-locality-diagram}}

\global\long\def\locgr{\mathcal{G}_{\mathtt{loc}}}%

Two primary fields of a conformal model are mutually local if their
OPE coefficients are single-valued functions of their separation.
Using conformal invariance, this translates into the requirement that,
denoting by $\cw p$ the conformal weight of a primary $p$, the primaries
$p$ and $q$ are mutually local if the difference $\cw r\!-\!\cw p\!-\!\cw q$
is an integer for any primary $r$ such that the fusion rule coefficient
$N_{pq}^{r}$ is positive.

Clearly, mutual locality of primaries is a symmetric binary relation,
which can be represented by an undirected graph (with possible loops),
whose vertices are the primary fields, with two of its vertices adjacent
whenever the corresponding primaries are mutually local. We shall
call this graph $\locgr$ the locality graph of the given conformal
model\footnote{Note that every locality graph is connected, since it contains a universal
vertex (i.e. one adjacent to all other vertices) corresponding to
the vacuum primary.}. As we shall see, it is the key to understanding the structure of
$\lat$, but to explain how this comes about, we have to take first
a look at some results about undirected graphs \citep{Balakrishnan1997,Bollobas2002}.

Given an undirected graph $\mathcal{\lgr}$, one can associate to
each vertex $v$ its neighborhood $\star v$, the collection of all
those vertices that are adjacent to it; in case of the locality graph
$\locgr$, the neighborhood of a primary will consist of those primaries
with which it is mutually local. More generally, to any collection
$X$ of vertices one can associate their common neighborhood $\star X$,
which consists of all vertices adjacent to each vertex in $X$, i.e.
the intersection of the neighborhoods of the vertices in $X$. It
is immediate that $X\!\subseteq\!Y$ implies $\star Y\!\subseteq\!\star X$,
and that $X\!\subseteq\!\cl X\!=\!\star{\star X}$. As a result, the
assignment $X\!\mapsto\!\cl X$ is a closure operation on sets of
vertices, hence the collection $\latg{\lgr}$, ordered by inclusion,
of those sets of vertices for which $\cl X\!=\!X$ (the closed ones)
is a finite lattice. What is more, this lattice $\latg{\lgr}$ comes
naturally equipped with a duality map, i.e. an involutive and order-reversing
self-map that assigns to each $X\!\in\!\latg{\lgr}$ the set $\star X\!\in\!\latg{\lgr}$;
put differently, the lattice $\latg{\lgr}$ is self-dual.

The basic result is that the lattice $\latg{\locgr}$ associated to
the locality graph coincides with the lattice $\lat$ of $\fc$s,
in such a way that the corresponding duality maps are equal. This
is indeed a truly remarkable fact, since it exhibits a close relation
between the fusion algebra and the locality graph, while neither of
these is completely fixed by the other. As a direct application, one can determine
the collection of all $\fc$s in a conformal model from the mere knowledge
of the locality graph, without having to deal with the details of
the fusion algebra. Even more is true, since the equality of the corresponding
duality maps allows to single out the local $\fc$s that form the
input of the deconstruction algorithm: these will correspond to those
closed sets $X\!\in\!\latg{\lgr}$ that are contained in $\star X$.
As a consequence, $\fc$s corresponding to maximal deconstructions
are in one-to-one correspondence with maximal cliques of the locality
graph $\locgr$ (more precisely, those maximal cliques all of whose
vertices are self-adjacent). Finding these is a classical problem
of graph theory \citep{Cormen2001}, with many applications ranging
from bioinformatics \citep{Day_1986} through electrical engineering
to social network analysis \citep{StanleyWasserman1995}.

While the above ideas already provide a serious improvement in the
handling of $\fc$s, they still have the drawback that, since the
size (i.e. number of vertices) of the locality graph equals the number of different primaries,
the computational cost of determining the corresponding lattice still
grows exponentially with the latter. But one can remedy this situation,
as we shall now explain.

Two vertices of an undirected graph $\lgr$ are said to be equilocal
if their neighborhoods coincide. Clearly, this is an equivalence relation,
whose equivalence classes partition the set of vertices in a way that is compatible
with adjacency. Put another way, the $\pcl$es provide a modular partition
of the graph \citep{Bollobas2002}. This allows to consider the quotient
graph $\defl{\lgr}$ of $\lgr$ by the equilocality relation\footnote{Note that this quotient graph is always irreducible in the sense that
no two of its vertices are equilocal.}, whose vertices correspond to equilocality classes, with two classes
being adjacent whenever they contain adjacent vertices. We shall call
the corresponding quotient $\defl{\locgr}$ of the locality graph
the locality diagram of the conformal model.

It follows from the symmetry of the adjacency relation that any element of $\latg{\lgr}$ is actually a union of equilocality classes.
As a result, there exists a map sending each $X\!\in\!\latg{\lgr}$
to the collection of equilocality classes that it contains, and one
may show that this deflation map induces an isomorphism between the
lattices $\latg{\lgr}$ and $\latg{\defl{\lgr}}$ that commutes with
the respective duality maps. It follows that the structure of the
lattice $\lat$ of $\fcs$s is completely determined by the collection
of $\pcl$es of the locality graph and by the lattice $\latg{\defl{\locgr}}$
associated to the locality diagram: as far as one is only interested
in the lattice structure of $\lat$, one may dispense with the locality
graph itself, and concentrate solely on its quotient, the locality
diagram.

This last result has many important consequences. In particular, it
leads to a dramatic decrease in the computational resources needed
to determine the lattice $\lat$, since there are usually much less
$\pcl$es than there are primary fields. This leads to the following effective procedure:
first, determine the  locality graph $\locgr$ of the model (the computational cost  being polynomial in the number of primaries),
from which one can read off the $\pcl$es and the locality diagram
$\defl{\locgr}$;
the next step (whose cost is exponential in the number of $\pcl$es) is to  compute the lattice $\latg{\defl{\locgr}}$ associated to the
 locality diagram;  finally,
one has to use the deflation isomorphism between $\latg{\defl{\locgr}}$ and
$\latg{\locgr}\!=\!\lat$ to get the result. This opens the way to
perform the necessary analysis for models with several hundreds, even
thousands of primaries.

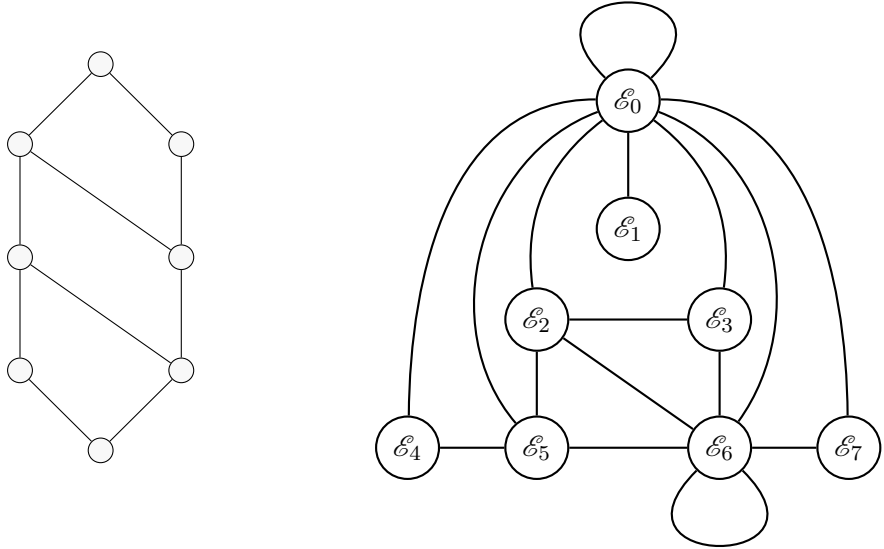
\begin{figure}[H]
~~~~~~~~~~~~~~~~~\begin{tikzpicture}[auto,node distance=1.5cm,
every loop/.style={}, 
null/.style={coordinate},
main node/.style={circle,draw,fill=gray!6}]
\node[main node] (8) {};  
\node[main node] (7) [below right of=8] {};    
\node[main node] (6) [below left of=8] {};  
\node[main node] (5) [below of=7] {};
\node[main node] (4) [below of=6] {};
\node[main node] (3) [below of=5] {};
\node[main node] (2) [below of=4] {};
\node[main node] (1) [below right of=2] {};
\node (0) [below of=1] {};
\path[every node/.style={font=\sffamily\small}]        
(8) edge node {} (7) 
edge node {} (6)  
(6)edge node [below right] {} (5)
edge node {} (4)
(7) edge node  {} (5)    
(5) edge  node {} (3)
(4) edge node {} (2)
edge node [below right] {} (3)
(3) edge node {} (1)
(2) edge node {} (1); 
\end{tikzpicture}~~~~~~~~~~~~~~~~~~~~~\begin{tikzpicture}[auto,node distance=1.7cm,
every loop/.style={}, thick,
null/.style={coordinate},
main node/.style={circle,draw}]
\node[main node] (1) {${\mathscr E}_1$};
\node [main node] (0) [above of=1] {${\mathscr E}_0$};  
\node[main node] (2) [below left of=1] {${\mathscr E}_2$};   
\node[main node] (3) [below of=2] {${\mathscr E}_5$};   
\node[main node] (4) [below right of=1] {${\mathscr E}_3$};
\node[main node] (5) [below of=4] {${\mathscr E}_6$};
\node[main node] (6) [left of=3] {${\mathscr E}_4$};
\node[main node] (7) [right of=5] {${\mathscr E}_7$};
\path[every node/.style={font=\sffamily\small}]        
(1) edge node {} (0)
(0) edge [loop] node {} (0)
edge [bend right] node {} (2) 
edge [bend left]  node {} (4) 
edge [bend right, out=305, in=235] node  {} (3)
edge [bend right, out=300] node  {} (6)
edge [bend left, out=55, in=130]  node {} (5)
edge [bend left, out=60]  node {} (7)
(2) edge node {} (3)
edge node {} (4)
edge node {} (5)
(3) edge node  {} (6)    
edge node {} (5)
(4) edge node  {} (5)  
(5) edge [in=315,out=225,loop] node {} (5)
edge node {} (7); 
\end{tikzpicture}\caption{\label{fig:Generic-Virasoro}Hasse diagram of the generic Virasoro
lattice $\protect\lat_{\mathtt{Vir}}$, and the corresponding locality
diagram (the labelling of the $\pcl$ being compatible with that of \prettyref{tab:Virasoropcl}).}
\end{figure}

Besides its computational utility, there is another, more conceptual
aspect of the deflation isomorphism: it does explain why so many,
at first sight pretty differently looking conformal models share the
same lattice $\lat$. This is due to the fact that, while their locality
graphs are truly different, the corresponding locality diagrams coincide
in many cases, leading to the same lattice structure. A nice example
of this phenomenon is provided by the (unitary) Virasoro minimal models:
while their structure is different, their lattice of $\fcs$s are,
except for two of them, all isomorphic to a generic Virasoro lattice
$\lat_{\mathtt{Vir}}$, whose Hasse diagram is depicted on the left
of \prettyref{fig:Generic-Virasoro}. The reason for this is that,
while the locality graphs differ from each other, the locality diagrams
are all isomorphic to the graph shown on the right of \prettyref{fig:Generic-Virasoro},
with the exception of the models with respective central charges $c\!=\!\nicefrac{7}{10}$
and $\nicefrac{1}{2}$, whose locality diagrams are shown in \prettyref{fig:lowVirasoro}.

Actually, even the two cases that do not fit in this generic pattern
may be understood as degenerations of the latter. Indeed, the corresponding
models have too few (only 6, resp. 3) primaries to fill each of the
8 $\pcl$es in \prettyref{fig:Generic-Virasoro}, so some of them
have to be left empty, and the corresponding vertices should left
out from the relevant diagram. The primary content of the different
$\pcl$es of Virasoro minimal models is summarized in \prettyref{tab:Virasoropcl},
showing that for the model with central charge $c\!=\!\nicefrac{7}{10}$
the $\pcl$es labeled $\mathscr{E}_{2}$ and $\mathscr{E}_{7}$ are
empty, hence the corresponding vertices are missing from the relevant
locality diagram, while for central charge $c\!=\!\nicefrac{1}{2}$
the classes labeled $\mathscr{E}_{1},\mathscr{E}_{3}$ and $\mathscr{E}_{5}$
are also void, so the relevant vertices have to be left out as well.
The resulting locality diagrams are depicted in \prettyref{fig:lowVirasoro}. 

\begin{table}[h]
\centering{}%
\begin{tabular}[b]{|c|c|c|c|c|}
\cline{2-5} \cline{3-5} \cline{4-5} \cline{5-5} 
\multicolumn{1}{c|}{} & ${\displaystyle c\!=\!\nicefrac{1}{2}}$ & ${\displaystyle c\!=\!\nicefrac{7}{10}}$ & ${\displaystyle c\!=\!\nicefrac{4}{5}}$ & ${\displaystyle c\!=\!\nicefrac{6}{7}}$\tabularnewline
\hline 
$\mathscr{E}_{0}$ & $\left(1,1\right)$ & $\left(1,1\right)$ & $\left(1,1\right)$ & $\left(1,1\right)$\tabularnewline
\hline 
$\mathscr{E}_{1}$ &  & $\left(2,2\right)$ & $\left(2,2\right),\left(3,2\right)$ & $\left(2,2\right),\left(4,2\right),\left(4,3\right),\left(4,4\right)$\tabularnewline
\hline 
$\mathscr{E}_{2}$ &  &  & $\left(4,3\right)$ & $\left(3,1\right)$\tabularnewline
\hline 
$\mathscr{E}_{3}$ &  & $\left(3,3\right)$ & $\left(2,1\right)$ & $\left(5,3\right),\left(5,5\right)$\tabularnewline
\hline 
$\mathscr{E}_{4}$ & $\left(2,2\right)$ & $\left(2,1\right)$ & $\left(4,2\right),\left(4,4\right)$ & $\left(2,1\right),\left(4,1\right)$\tabularnewline
\hline 
$\mathscr{E}_{5}$ &  & $\left(3,2\right)$ & $\left(3,1\right)$ & $\left(5,2\right),\left(5,4\right)$\tabularnewline
\hline 
$\mathscr{E}_{6}$ & $\left(2,1\right)$ & $\left(3,1\right)$ & $\left(4,1\right)$ & $\left(5,1\right)$\tabularnewline
\hline 
$\mathscr{E}_{7}$ &  &  & $\left(3,3\right)$ & $\left(3,2\right),\left(3,3\right)$\tabularnewline
\hline 
\end{tabular}\caption{\label{tab:Virasoropcl}Kac labels of the primaries filling the different
equilocality classes of the first few Virasoro minimal models.}
\end{table}
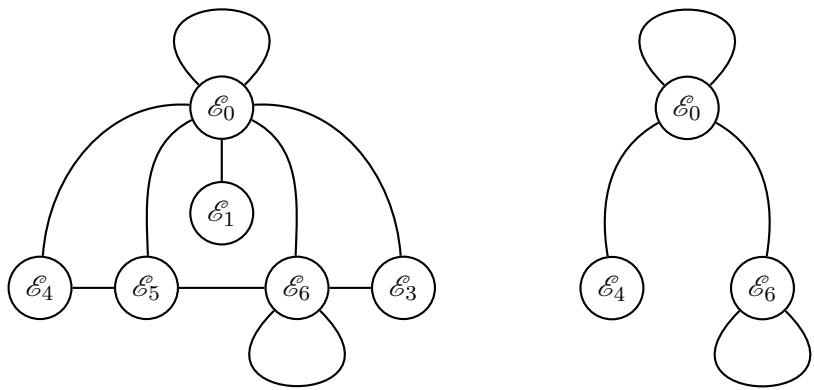
\begin{figure}[h]
~~~~~~~~~~~~~~~~~~~\begin{tikzpicture}[auto,node distance=1.4cm,
every loop/.style={}, thick,
ess/.style={circle,draw},
main node/.style={circle,draw}]
\node[main node] (1) {${\mathscr E}_1$};
\node [ess] (0) [above of=1] {${\mathscr E}_0$};     
\node[ess] (4) [below left of=1] {${\mathscr E}_5$};
\node[ess] (5) [below right of=1] {${\mathscr E}_6$};  
\node[main node] (3) [right of=5] {${\mathscr E}_3$};
\node[ess] (6) [left of=4] {${\mathscr E}_4$};
\path[every node/.style={font=\sffamily\small}]        
(1) edge node {} (0)
(0) edge [loop] node {} (0)
edge [bend right, out=315, in=200] node  {} (4)
edge [bend right, out=310, in=220] node  {} (6)
edge [bend left, out=45, in=160]  node {} (5)
edge [bend left, out=50, in=140]  node {} (3)
(3) edge node {} (5)
(4) edge node  {} (5) 
edge node {} (6)
(5) edge [in=315,out=225,loop] node {} (5); 
\end{tikzpicture}~~~~~~~~~~~~~~~\begin{tikzpicture}[auto,node distance=1.4cm,
every loop/.style={}, thick,
ess/.style={circle,draw},
main/.style={circle}]
\node[ess] (1) {${\mathscr E}_0$};
\node[main] (3) [below of=1] {};
\node [ess] (0) [below left of=3] {${\mathscr E}_4$};   
\node[ess] (2) [below right of=3] {${\mathscr E}_6$};
\path[every node/.style={font=\sffamily\small}]        
(1) edge [loop] node {} (1)
edge [bend right, out=320, in=210] node {} (0)
edge [bend left, out=40, in=150] node {} (2)
(2) edge [in=315,out=225,loop] node {} (2);
\end{tikzpicture}\caption{\label{fig:lowVirasoro}Locality diagrams of the Virasoro minimal
models of respective central charges $c\!=\!\nicefrac{7}{10}$ and
$\nicefrac{1}{2}$.}
\end{figure}

\begin{figure}[t]
\begin{centering}
~~~~~~\usetikzlibrary {shapes.geometric} 
\begin{tikzpicture}[auto,node distance=1.4cm,
every loop/.style={}, thick,
ess/.style={circle,draw,fill=gray!6},
null/.style={coordinate},
main/.style={circle,draw,fill=gray!6}]
\node[ess] (0) {}; 
\node[main] (1) [above of=0] {};
\node[null] (15) [above of=1] {};
\node[null] (12) [below left of=0] {};
\node[ess] (3) [left of=12] {};
\node[main] (2) [left  of=3] {};
\node[null] (13) [below right of=0] {};
\node[ess] (4) [right of=13] {};
\node[main] (5) [right of=4] {};
\node[ess] (6)  [below left of=13] {};
\node[main] (7)  [below  of=6] {};
\node[null] (14) [below right of=7] {};
\node[ess] (8)  [below  of=7] {};
\node[main] (9)  [below right of=6] {};
\node[main] (10)  [below left of=8] {}; 
\node[main] (11)  [below right of=8] {}; 
\path[every node/.style={font=\sffamily\small}]        
(0) edge [bend right, out=305, in=235] node {} (2)
edge [bend right, out=310, in=225] node {} (3)
edge [bend right, out=315] node {} (10)
edge [bend left, out=45] node {} (11)
edge node {} (1)
edge [bend left, out=50, in=135] node {} (4)
edge [bend left, out=55, in=125] node {} (5)
edge [bend right, out=320] node {} (8)
edge  node {} (6)
edge [bend left, out=15] node {} (9)
edge [bend right, out=340] node {} (7)
edge [out=135, in=180] node {} (15)
(15) edge [out=0, in=45] node {} (0)
(2) edge node {} (3)
edge [out=335, in=170] node {} (6)
edge [out=305, in=155] node {} (8)
edge [out=285, in=135] node {} (10) 
edge [out=345,in=195] node {} (5)
(4) edge node {} (5)
(5) edge [out=205, in=10] node {} (6)
edge [out=235, in=25] node {} (8)
edge [out=255, in=45] node {} (11)
(6) edge node {} (7)
(7) edge node {} (8)
(8) edge node {} (9)
edge node {} (10)
edge node {} (11)
edge [in=305,out=235,loop] node {} (8)
(10) edge [out=345, in=195] node {} (11); 
\end{tikzpicture}
\par\end{centering}
\centering{}\caption{\label{fig:trilobite}Locality diagram of $N\!=\!2$ superconformal
minimal models of central charge $c\!=\!3\!-\!\frac{6}{p}$, with
$p$ an odd prime.}
\end{figure}
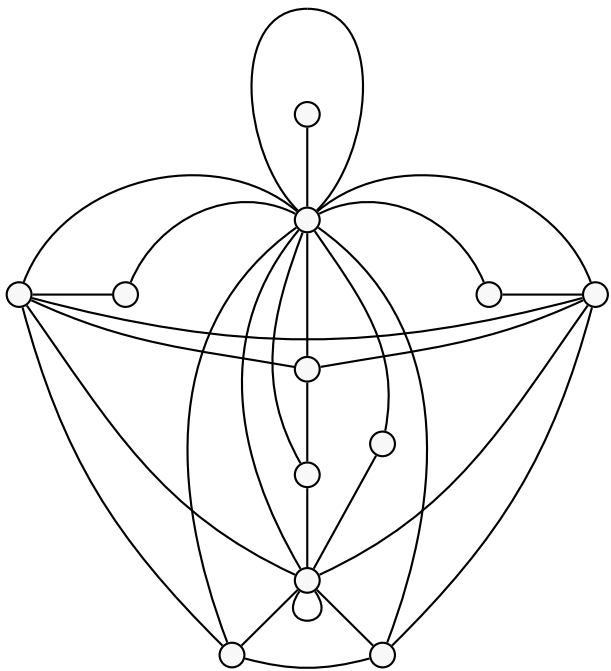

Similar results hold for other classes of conformal models, like Wess-Zumino,
superconformal, parafermionic, etc., but usually the pattern is more
complicated, with several classes of 'generic' diagrams and their
different degenerations. In case of $N\!=\!2$ superconformal minimal
models, the locality diagram for models of central charge $c$ seems
to be determined by the primary decomposition (as a product of prime
powers) of the integer $\frac{6}{3-c}$, the trilobite-like diagram
depicted in \prettyref{fig:trilobite} corresponding to the case when
this last number is actually an odd prime. Similar patterns can be
observed for parafermionic and Ashkin-Teller models (i.e. $\mathbb{Z}_{2}$
orbifolds of a free boson compactified on a circle of suitable radius).

Primaries that belong to the same $\pcl$ share many properties. For example, if the quantum dimension of a primary equals
$1$ (resp. is an integer), then the same is true for all primaries
in the same $\pcl$. More generally, one may show that the number
field generated by the quantum dimension of a primary is the same
for all elements of its $\pcl$. In particular, self-local (i.e. self-adjacent)
$\pcl$es are integral in the sense that the quantum dimension of
all their elements are rational integers, and their conformal weights
are either integers or half-integers.

An interesting aspect of $\pcl$es is related to orbifold
deconstruction. Indeed, consider a twister $\mathfrak{g}$ corresponding
to a realization of the given conformal model as an orbifold. According to  general principles of orbifold deconstruction
\citep{Bantay2019a,Bantay2020}, the elements of the twister $\mathfrak{g}$
correspond to irreducible representations of the twist group, and
two elements of $\mathfrak{g}$ are equilocal precisely when the kernels
of the associated representations coincide. This means that to each
$\pcl$ contained in a twister is associated a normal subgroup of
the corresponding twist group. Conversely, to each normal subgroup
 of the twist group  there corresponds a partial deconstruction
whose vacuum block $\mathfrak{h}$ is a twister contained in $\mathfrak{g}$,
and the normal subgroup  can be recovered as the intersection
of the normal subgroups associated to the different $\pcl$es contained
in $\mathfrak{h}$.

\begin{figure}[h]
~~~~~~~~~~~~~~~~~~~~~~~~~~\begin{tikzpicture}[auto,node distance=1.2cm,
every loop/.style={}, 
main node/.style={circle,draw,fill=gray!6},
null/.style={coordinate}]
\node[main node] (8) {};   
\node[main node] (7) [below right of=8] {};  
\node[null] (4)  [below left of=7] {};
\node[main node] (6) [left of=4] {};  
\node[main node] (5) [below right of=4] {};
\node[main node] (1) [below left of=5] {};
\path[every node/.style={font=\sffamily\small}]        
(8) edge node {} (7) 
edge node {} (6)  
(6)edge node {} (1)
(7) edge node  {} (5)    
(5) edge  node {} (1); 
\end{tikzpicture}~~~~~~~~~~~~~\begin{tikzpicture}[auto,node distance=2.0cm,
every loop/.style={}, thick,
main node/.style={circle,draw}]
\node[main node] (0) [] {};   
\node[main node] (1) [below of=0] {};
\node[main node] (2) [left of=1] {};
\node[main node] (3) [right of=1] {};
\path[every node/.style={font=\sffamily\small}]        
(0) edge [loop] node {} (0)
edge  node {} (2) 
edge   node {} (3) 
edge  node {} (1)
(1) edge node {} (2)
(2) edge node {} (1)
edge [out=165, in=265, loop] node {} (2)
(3) edge [in=275,out=15,loop] node {} (3); 
\end{tikzpicture}\caption{\label{fig:nonmodular}Hasse diagram of a non-modular lattice, and
the corresponding graph.}
\end{figure}
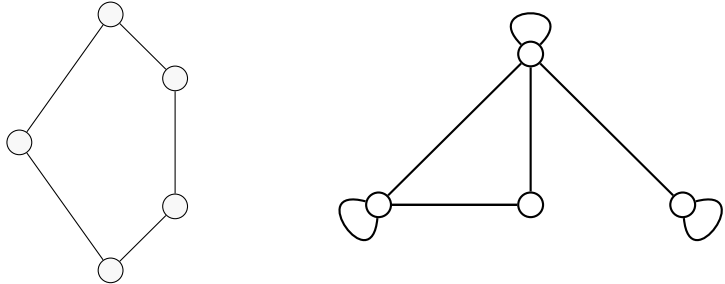

While locality diagrams are helpful in describing the lattice of $\fc$s,
we should note that not every irreducible graph is the locality diagram
of some conformal model. An obvious property follows from the existence
of the vacuum primary, which is mutually local with all the primaries,
and forms an $\pcl$ in itself, hence every locality diagram has a
universal vertex that is adjacent to all vertices (including itself).
A more subtle requirement comes from the lattice $\lat$ admitting
a type I embedding into a partition lattice, since there are undirected
graphs for which the associated lattice is not even modular, an example
being shown in \prettyref{fig:nonmodular}. 

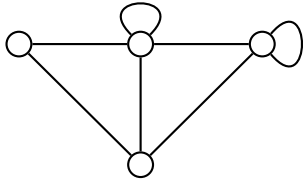
\begin{figure*}[h]
~~~~~~~~~~~~~~~~~~~~~~~~~~~~~~~~~~~~~~~~~~~~~~~~\begin{tikzpicture}[auto,node distance=1.6cm,
every loop/.style={}, thick,
main node/.style={circle,draw}]
\node[main node] (0) [] {};   
\node[main node] (1) [below of=0] {}; 
\node[main node] (2) [left of=0] {};
\node[main node] (3) [right of=0] {};
\path[every node/.style={font=\sffamily\small}]        
(0) edge [loop] node {} (0) 
edge  node {} (3) 
edge  node {} (2)
edge  node {} (1)
(3) edge [out=50, in=310, loop] node {} (3)
edge node {} (1)
(2) edge node {} (1); 
\end{tikzpicture}\caption{\label{fig:nonweighted}An irreducible graph that does not correspond
to a locality diagram, while its associated lattice is modular.}
\end{figure*}

Actually, the situation is a bit more complicated, as there are unoriented
graphs that satisfy all the above criteria, namely that they are irreducible,
have a universal vertex, and their associated lattice admits a type
I embedding into a partition lattice, but cannot show up as the locality
diagram of a conformal model. The reason can be traced back to the
fact mentioned at the end of \prettyref{sec:FC-sets-and}, namely
that the product $\sp{\mathfrak{g}}\!\sp{\du{\mathfrak{g}}}$ is the
same for every $\fc$ $\mathfrak{g}\!\in\!\lat$. Taking into account
Eq.\eqref{eq:sprmod}, this requirement can be translated into a set
of quadratic equations to be satisfied by the values assigned to the
different $\pcl$es, and these equations must have a positive solution
for the graph to be realizable as the locality diagram of some conformal
model. That this is not automatic is illustrated by the graph shown
on \prettyref{fig:nonweighted}, whose associated lattice is identical
to that of the Virasoro minimal model of central charge $c\!=\!\nicefrac{7}{10}$,
but nevertheless cannot be the locality diagram of a conformal model
because it does not satisfy this last requirement.

\section{Summary}

As we tried to explain in the previous sections, the full deconstruction
hierarchy, and the closely related lattice $\lat$ of $\fc$s can
be determined from the sole knowledge of the locality graph, or even
better, from the locality diagram and the primary content of the individual
equilocality classes. Not only does this simplify dramatically actual
computations, but it does also provide new insights into the structure
of conformal models, and in particular, it brings to the forefront
the notion of $\pcl$es, and the relationship between primaries in
the same class. 

Actually, the existence of generic locality diagrams makes it possible
to compare not only primaries of a given model, but also primaries
coming from different models (provided the relevant locality diagrams are isomorphic,
or at least degenerations of a common graph), establishing some kind of 'kinship'
between them. For example, the $\pcl$
labeled $\mathscr{E}_{6}$ on \prettyref{fig:Generic-Virasoro} is
present in all Virasoro minimal models, and it contains a single primary
field having analogous properties in each of them. Such relations
between primaries of different (although somehow related) models could
prove helpful, e.g. in classification attempts.

Finally, it should be pointed out that, while our exposition was formulated
using notions of 2D CFT \citep{BPZ,DiFrancesco-Mathieu-Senechal},
most of the ideas and results presented could be directly applied
to such related fields as the theory of Vertex Operator Algebras \citep{Borcherds1,FLM1,Kac}
or that of Modular Tensor Categories \citep{Turaev,Bakalov-Kirillov},
providing potential new insights. In particular, one may speculate
about their application in the analysis of topological order \citep{Wen2007}.
From a more general perspective, application of techniques from graph
theory to the study of QFT seems a most interesting possibility.

\bibliographystyle{plain}

\end{document}